# Disease Insight through Digital Biomarkers Developed by Remotely Collected Wearables and Smartphone Data

**Zulqarnain Rashid[1], PhD; Amos A Folarin[12345], PhD; Yatharth Ranjan[1], MSc; Pauline Conde[1], BSc; Heet Sankesara[1], BSc; Yuezhou Zhang[1], PhD; Shaoxiong Sun[1], PhD; Callum Stewart[1], PhD; Petroula Laiou[1], PhD; Richard JB Dobson[12345], PhD;**

1 Institute of Psychiatry, Psychology and Neuroscience, King's College London.
2 Institute of Health Informatics, University College London, London, United Kingdom.
3 NIHR Biomedical Research Center at South London and Maudsley NHS Foundation Trust and King's College London, London, United Kingdom.
4 Health Data Research UK London, University College London, London, United Kingdom.
5 NIHR Biomedical Research Center at University College London Hospitals, NHS Foundation Trust, London, United Kingdom.

## Corresponding Authors

**Amos A Folarin, Richard JB Dobson**
Institute of Psychiatry, Psychology and Neuroscience
King's College London
SGDP Centre, IoPPN, Box PO 92
De Crespigny Park, Denmark Hill, London , SE5 8AF
Tel: +44(0)207 848 0924
email: (amos.folarin, richard.j.dobson)@kcl.ac.uk

## Abstract

**Background**
Digital Biomarkers and remote patient monitoring can provide valuable and timely insights into how a patient is coping with their condition (disease progression, treatment response, etc.), complementing treatment in traditional healthcare settings.

**Objectives**
Smartphones with embedded and connected sensors have immense potential for improving healthcare through various apps and mHealth (mobile health) platforms. This capability could enable the development of reliable digital biomarkers from long-term longitudinal data collected remotely from patients.

**Methods**
We built an open-source platform, RADAR-base, to support large-scale data collection in remote monitoring studies. RADAR-base is a modern remote data collection platform built around Confluent's Apache Kafka, to support scalability, extensibility, security, privacy and quality of data. It provides support for study design and set-up, active (eg PROMs) and passive (eg. phone sensors, wearable devices and IoT) remote data collection capabilities with feature generation (eg. behavioural, environmental and physiological markers). The backend enables secure data transmission, and scalable solutions for data storage, management and data access.

**Results**
The platform has successfully collected longitudinal data for various cohorts in a number of disease areas including Multiple Sclerosis, Depression, Epilepsy, ADHD, Alzheimer, Autism and Lung diseases. Digital biomarkers developed through collected data are providing useful insights into different diseases.

**Conclusion**
RADAR-base provides a modern open-source, community-driven solution for remote monitoring, data collection, and digital phenotyping of physical and mental health diseases. Clinicians can use digital biomarkers to augment their decision making for the prevention, personalisation and early intervention of disease.

## Introduction

Digital biomarkers offer a host of advantages to measure our health over traditional biomarkers that are typically confined to clinical settings, including decentralisation, scalability, sampling frequency and real-time measurement, and affordability. However, significant challenges remain with implementing digital biomarkers.

Digital biomarkers developed from sensor data can help with preventions and early intervention to better diagnose and manage disease. Collected data should be reliable and of high quality reflecting the true condition of the patients and many studies have attempted to measure the effectiveness of digital biomarkers for various clinical use cases (1).

For reliable digital biomarker development long term longitudinal high quality data of digital sensors is required. The widespread availability of smartphones, more capacious mobile networks and the development of new wearable sensors has enabled measurement of a growing set of physiological and phenomenological parameters relevant to physical and mental diseases. To facilitate the wearables and smartphone data remote collection at scale and digital biomarkers development, the RADAR-base platform was released under the open-source Apache 2 licence in January 2018 (2) (3). RADAR-base is composed of an Apache Kafka-based back-end deployed onto Kubernetes infrastructure and two mobile apps. The cross-platform (Android, iOS) Cordova Active Remote Monitoring App (aRMT) for active monitoring of participants , requiring conscious action (e.g. questionnaires, audio questions, timed tests etc.), and the Passive Remote Monitoring App (pRMT), a native Android app for passive monitoring of participants via phone, wearable devices and IoT. The high-level overview of the platform is shown in Figure 1.

The aRMT app provides highly extensible active remote monitoring functionality to the platform, rendering questionnaires from a JSON definition file and a schedule defined by a protocol.json file which determines the ordering and notification regimen and some additional configuration metadata. Some questionnaires used with the aRMT app in projects include RSES, PHQ8 and ESM, voice/audio sampling, and clinical tests (Rhomberg and walking tests).

The passive application runs in the background, requiring minimal or no input from participants. Data is collected from smartphone "sensors" and from integrated wearable devices. The catalogue of devices currently integrated into the pRMT app includes onboard Android smartphone sensors, Empatica E4, Pebble 2 smartwatch, Biovotion Everion, Faros 180 and 360, Fitbit, Garmin Vivosmart, Oura Ring. Pluggable capability is provided to integrate new wearable devices offering a native SDK (e.g. Empatica E4) or through 3rd party vendor's REST API (e.g. Fitbit via the backend REST Collector + REST-Authorizer for OAuth-2 Flows).

A common task is the exploration of collected raw data. RADAR-base includes capability for data aggregation, management of studies, and real time visualisation in Grafana dashboards (4). In addition to the near real-time visualisation provided by the dashboard, the RADAR-base platform includes a Python package designed for data processing, feature generation and visualisation. This package offers a range of standard tools for exploratory visualisations of collected data. It also simplifies the implementation of feature generation pipelines, allowing users to take data exported from a RADAR-base project and generate processed data (high-level features), along with any associated labels, in a format suitable for use with commonly used machine learning libraries.

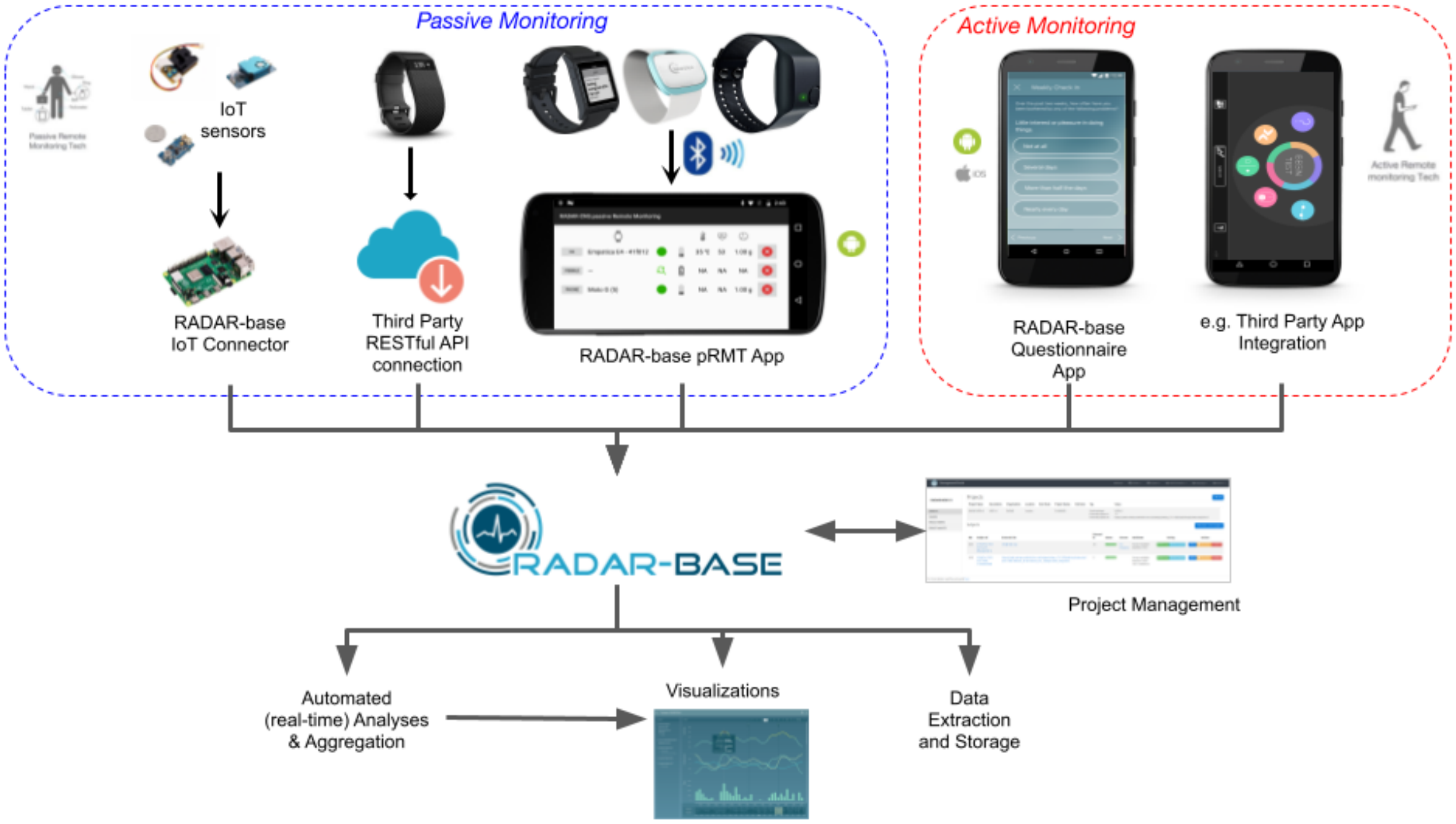

Figure 1. Overview of the RADAR-base Platform. Current data sources: Empatica E4, Pebble 2, Fitbit, Biovotion, Faros, Garmin; Active aRMT Questionnaire app, and Passive pRMT app.

The Platform is used in a wide range of research and clinical studies and is available under a range of service models, depending on the requirement of the project. This paper will discuss and summarise a representative selection of the research and clinical studies using the RADAR-base platform for data collection and digital phenotyping, focusing on common patterns of use and challenges.

## Related Work

Numerous studies are validating digital biomarkers for disorders and their effectiveness, a study (5) was conducted to assess the usefulness of digital biomarkers for mood and depression, however it was a small cohort of only 60 participants. Digital biomarkers are being studied with a view to replacing, or augmenting traditional markers for disorders and a number of barriers have been identified, these barriers include standardisation and regulation, studies are undergoing to address challenges and streamline digital biomarkers in healthcare (6). Open source software platform for end-to-end digital biomarker development Digital Biomarker Discovery Pipeline (DBDP) was developed to standardise digital biomarker development. DBDP modules calculate and utilise resting heart rate (RHR), glycemic variability, insulin sensitivity status, exercise response, inflammation, heart rate variability, activity, sleep, and circadian patterns to predict health outcomes (7). mCerebrum: A Mobile Sensing Software Platform for Development and Validation of Digital Biomarkers and Interventions supports high-rate data collections from multiple sensors with real time assessment of data quality and development of digital biomarkers (8).

Intel Context Sensing SDK is a library for Android and Windows with specific context states, it however only provides front-end components (9). The EmotionSense app is developed by the University of Cambridge to sense emotions with implications for psychological therapy and improving well-being, however, it is only focused on depression (10). Medopad provides solutions for different healthcare issues with symptom tracking, this is a commercial solution and mainly focuses on phone sensors and active monitoring methods (11). PHIT allows users to build health apps based on existing infrastructure (12). ResearchKit, an open-source framework for building apps specifically for iOS, ResearchKit makes it easier to enrol participants and conduct studies, however, new wearable device integration requires strong programming skills and it does not include a data management solution (13).

ResearchStack is an SDK and UX framework for building research study apps on Android, with a similar application domain as ResearchKit (14) Both ResearchKit and ResearchStack provide software libraries, frameworks, and development tools that require extensive programming skills to create apps. A framework to create observational medical studies for mobile devices without extensive programming skills was presented (15).

There is a need for a systematic approach to assess the quality and utility of digital biomarkers to ensure an appropriate balance between their safety and effectiveness. Study (16) outlines key considerations for the development and evaluation of digital biomarkers, examining their role in clinical research and routine patient care . RADAR-base provides a safe and effective digital biomarker ecosystem ensuring transparency of the algorithms, interoperable components with open interfaces to accelerate the development of new multicomponent systems, and high integrity measurement systems.

## Digital Phenotyping of Disease

A key feature of RADAR-base is its extensibility allowing new wearables and sensors to be readily integrated into the platform to collect new modes of data depending on the study requirements. Collected raw data from phone and wearable sensors can then be aggregated and converted into low level features, and subsequently high-level features, representing digital biomarkers . Figure 2 exemplifies the process for Major Depressive Disorder (MDD) for wearable and phone sensor collected data. E.g. phone microphone collected audio data can provide different speech features and respiratory acoustics which could help identify respiration and physiological stress, similarly low level acceleration provided actigraphy features can be used to identify psychomotor retardation. Developed digital biomarkers may provide insight into the disorder and could be used in clinical trials to ascertain their usefulness in management of the disease.

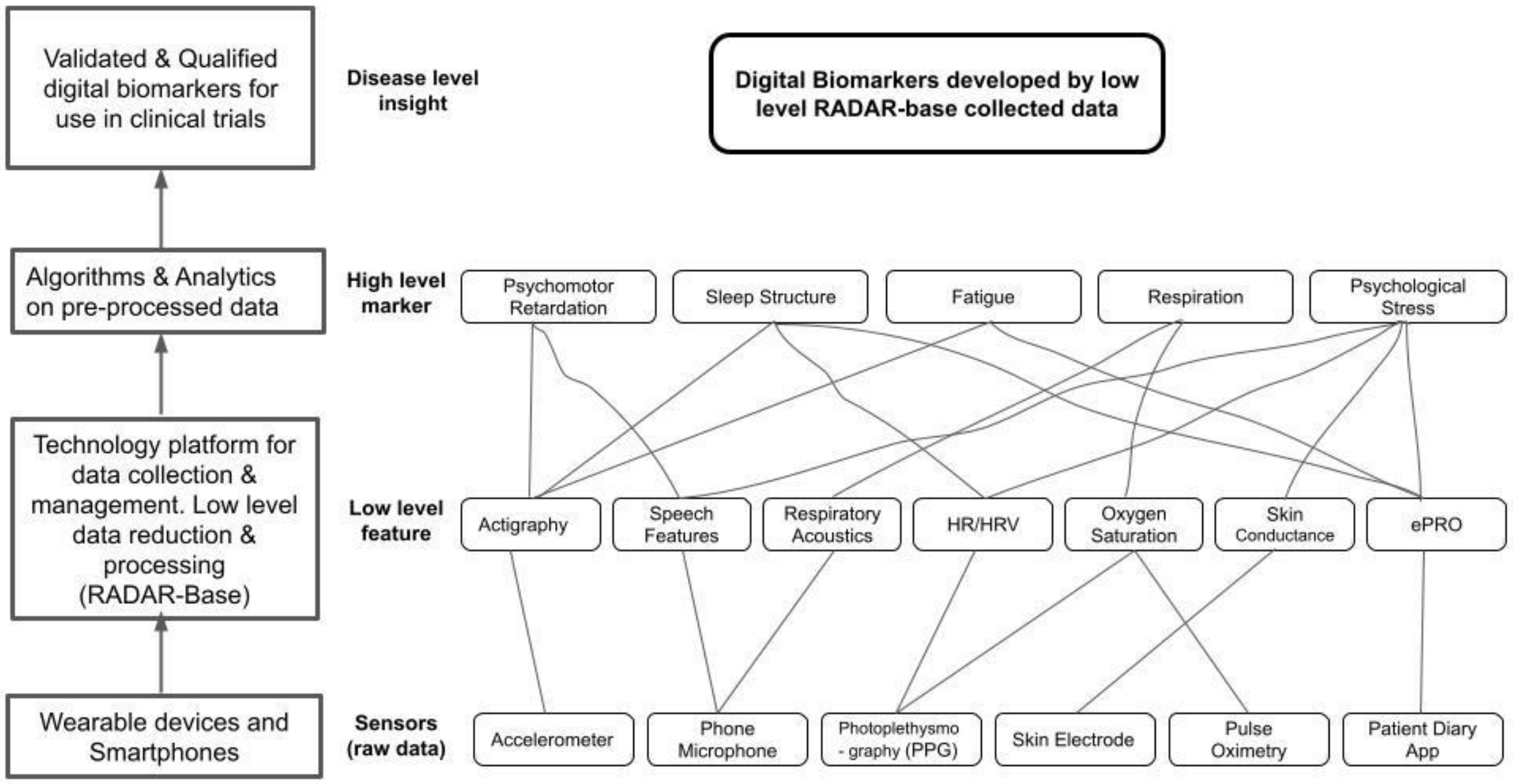

Figure 2. Collected raw data transition into high level features which gives insight into Major Depressive Disorder.

The next two sections discuss the development of high level features for MDD and Epilepsy and how these features are being used to explore and manage disease. Table 1 shows selected features extracted from sensor data and Table 2 presents digital biomarkers developed for different disorder areas (17) (18) (19). An open-source feature generation pipeline has been created to enhance and standardise the analysis of data generated by RADAR-base. This pipeline facilitates the extraction of features and biomarkers, enabling cross-disease symptom analysis. With its capabilities to ingest, analyse, visualise, and export RADAR-base data, the pipeline simplifies and establishes a convention for data scientists. This streamlines the process of feature-based analysis, ensuring consistency and enabling researchers to gain valuable insights from the data. Pipelines are readily extended and published on the RADAR-base pipeline catalogue (20).

### Major Depressive Disorder

Major Depressive Disorder (MDD) is associated with a wide range of negative outcomes including: premature mortality, reduced quality-of-life, loss of occupational function and is often experienced alongside physical comorbidity and approximately 55% will go on to develop chronic depression, characterised by periods of recovery and relapse (21).

The pRMT app provides a comprehensive solution for activity monitoring by utilising wearable device sensors and smartphones to collect data without requiring any input from the wearer. It leverages a range of sensors, including Global Positioning System (GPS), accelerometer, gyroscope, communication logs, ambient noise and light levels, and screen interactions. Through this approach, the app can effectively and passively gather diverse data streams, enabling a seamless and unintrusive data collection process for various applications. These sensors along with the Fitbit watch have the potential to identify changes in sleep, communication and activity patterns associated with depressive episodes. The smartphone embedded Bluetooth sensor can be used to record individuals' local proximity information, such as the nearby Bluetooth device count (NBDC) that includes the Bluetooth signal of other phone users. The NBDC data have the potential to reflect changes in people's behaviours and statuses during the depressive state (22). An illustration is given in Figure 3.

Speech characteristics, such as speaking rate, pitch, pause duration and energy, collected via smartphone microphones, have been used to detect depression with a prediction accuracy of 81.3% (23). aRMT app delivers validated questionnaires, cognitive games, speech tasks or electronic diaries using the experience sampling method (ESM) to provide fine grained understanding of mood changes and stressors in the context of daily life. The aRMT app has been used to measure affect, cognition, mood and behaviour in real time, with evidence highlighting the increased validity of this methodology in comparison to traditional retrospective reports. The aRMT assessments of positive and negative affect have also been found to be reliably indicative of mood state and have been associated with MDD symptoms (24).

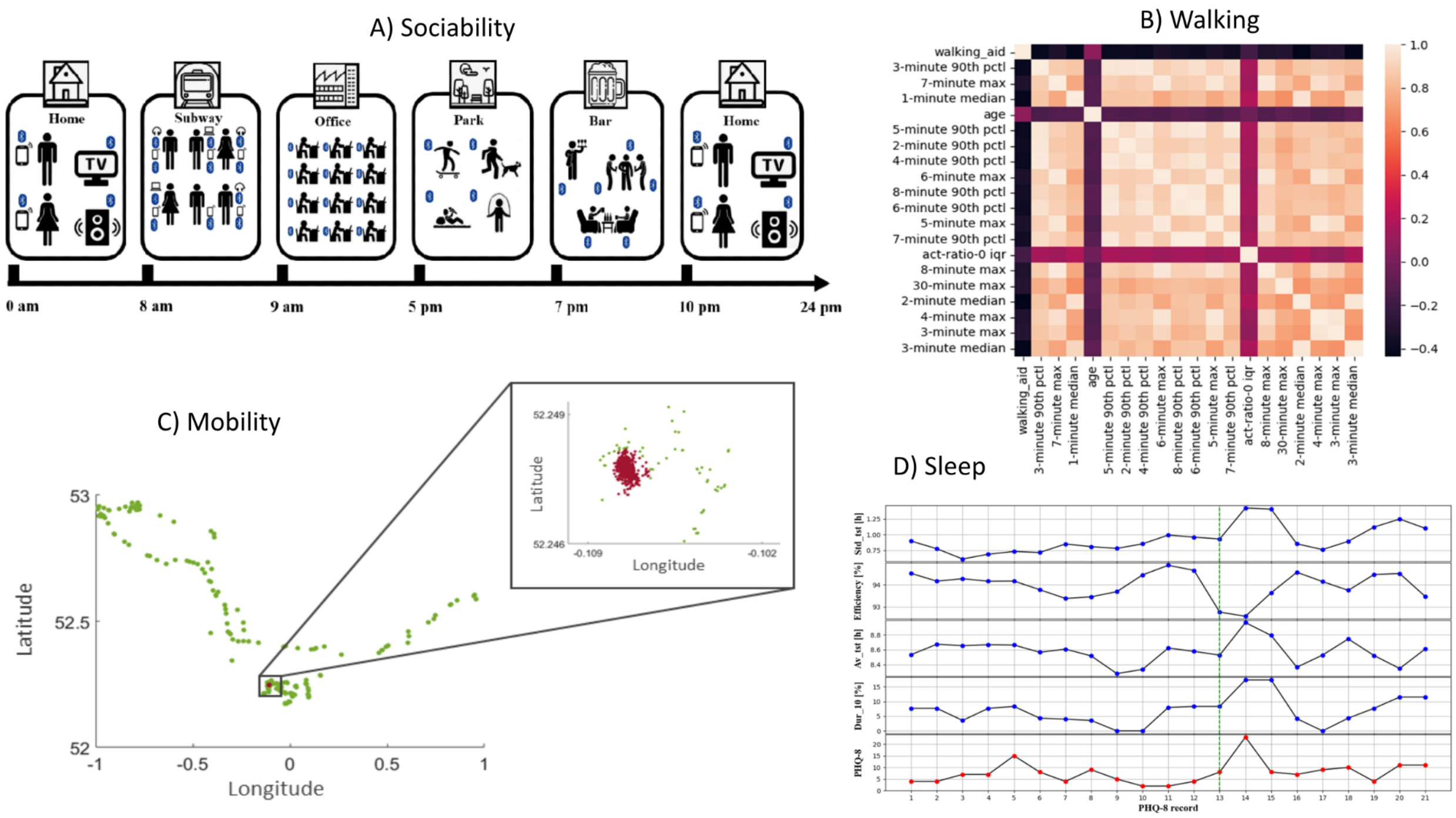

Figure 3.(a) A schematic diagram showing an individual's nearby Bluetooth devices count (NBDC) in different scenarios in daily activities and life. (b) Pearson correlation heatmap for top 20 features of walking data (median rankings from all models). (c) Exemplar geolocation data correspond to a biweekly segment of a study participant, The red dots denote an individual's home location, whereas longitude and latitude along the axes are expressed in decimal degrees. (d) The PHQ-8 scores and a select 4 sleep features of one participant with an obvious increasing trend in PHQ-8 score at 13th PHQ-8 record.

## Epilepsy

Numerous epilepsy research studies based on epilepsy monitoring units (EMUs), have shown the possibility of capturing characteristic movement associated with myoclonic seizure manifestations using wearable sensors (25). Using pRMT app integrated wearable sensors, it is possible to record several signals associated with seizure including motor components, using inertial sensors such as accelerometry and surface electromyography, various features of Heart rate variations captured by wearable electrocardiogram (ECG) and photoplethysmography (PPG), and alteration of the autonomic nervous system with ECG, PPG, and electrodermal activity (EDA) sensors, with different levels of signal and seizure detection accuracy. Figure 4 shows seizures detected with the collected data using the Empatica E4 wearable.

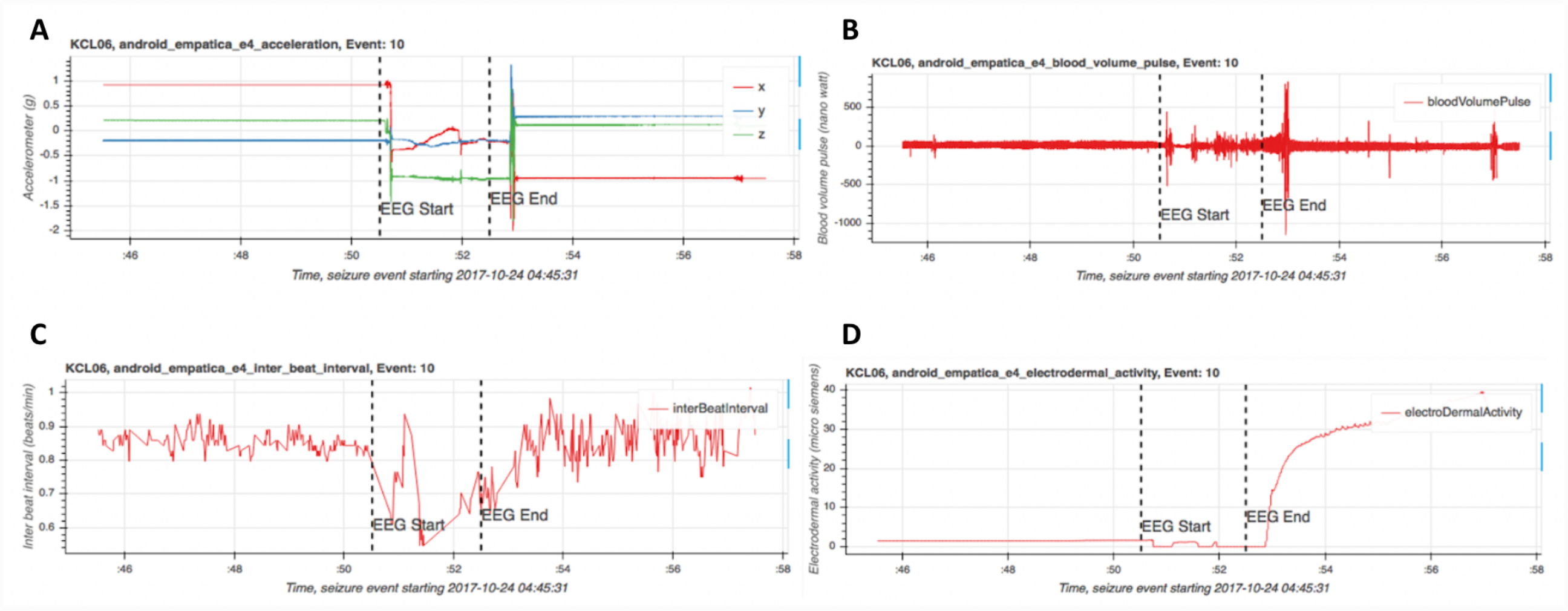

Figure 4. Empatica E4 sensor data. The area between the vertical dashed lines indicates a focal seizure with a motor component. (a) Accelerometer, (b) PPG blood volume pulse, (c) PPG inter-beat-interval, (d) Electrodermal activity.

Table1. Features extracted from the sensors integrated into the RADAR-base platform.

| Apps | Devices | Sensors | Selected Features |
|---|---|---|---|
| aRMT<br>pRMT app | Empatica E4<br>Biovotion<br>Everion<br>Bittium<br>Faros 180<br>Faros 360<br>Fitbit<br>Garmin<br>Smart Phone<br>Oura Ring | Acceleration<br>Blood Volume Pulse<br>Electrodermal<br>Activity<br>Inter Beat Interval<br>Temperature<br>Blood Pulse Wave<br>Galvanic Skin<br>Response<br>Heart Rate<br>Oxygen Saturation<br>Led Current<br>PPPG Raw<br>ECG<br>Gyroscope<br>Light<br>Magnetic Field<br>Relative Location<br>Step Count<br>Usage Event<br>User Interaction<br>Activity Levels | Sleep Duration<br>Sleep Architecture<br>Sleep Stability<br>Sleep Quality<br>Sleep Efficiency<br>Sleep Fragmentation Index<br>Sleep Onset Latency<br>Sleep Onsent Latency Variance<br>Sleep Midpoint<br>Sleep Midpoint Variance<br>Insomnia<br>Hypersomnia<br>Unlocktimes/Duration<br>Unlock Duration Min/Max<br>Median Interval Between Two<br>Unlocks<br>Step Epoch<br>Daily Step Sum<br>Moderate walking duration<br>Maximum non-stop duration<br>Maximum non-stop step count<br>Activity level |
| | | Activity Log Record<br>Intraday Calories<br>Intraday Steps<br>Resting Heart Rate<br>Sleep Classic<br>Sleep Stage<br>SMS Unread<br>Bluetooth Devices<br>Phone Battery Level<br>Phone Contact List<br>Garmin Stress Tracking<br>Garmin Relaxation Breathing timer<br>Garmin Vo2 max<br>Garmin Body Battery Energy Monitor<br>SpO2 | Actigraphy<br>Respiratory Acoustics<br>Heart Rate<br>Heart Rate Variation<br>Oxygen Saturation<br>Skin Conductance<br>ePRO<br>Ambient Light<br>Activity<br>Phone Use<br>Bluetooth<br>Max/Min/Mean/SD of nearby Bluetooth Devices (NBDC)<br>NBDC Entropy<br>NBDC Frequency Features<br>Location<br>Location variance<br>Moving time<br>Moving distance<br>Number of Location Clusters<br>Location Entropy<br>Homestay<br>Location Frequency Features<br>Gait<br>Median Gait Cycles<br>Frequency of gait<br>Median Force<br>Change in Total Sleep<br>Social Jet Lag |

Table 2. Digital biomarkers generated from extracted features, and associated studies.

| **Digital Biomarker** | RADAR-CNS | RADAR-AD | AIMS-2-TRIALS | ART | BigData@Heart | RALPMH | COVID-Collab |
|---|---|---|---|---|---|---|---|
| Total Sleep | ✔ | ✔ | ✔ | ✔ | ✔ | ✔ | ✔ |
| Social interactions | ✔ | ✔ | ✔ | ✔ | | | ✔ |
| Gait patterns | ✔ | ✔ | | | | | |
| Respiration | ✔ | ✔ | | | | ✔ | ✔ |
| Psychomotor retardation | ✔ | ✔ | | | | | |
| Psychological Stress | ✔ | | ✔ | ✔ | ✔ | ✔ | ✔ |
| Phone Use | ✔ | ✔ | ✔ | ✔ | ✔ | | |
| Ambulatory Mobility | ✔ | | ✔ | | | ✔ | |
| Fatigue | ✔ | | | | | ✔ | |
| Seizures | ✔ | | | | | | |
| Behavior | ✔ | ✔ | ✔ | ✔ | | | |
| Activity | ✔ | ✔ | ✔ | | | ✔ | |
| Speech | ✔ | | | | | | |
| Resting HR | ✔ | | | | | ✔ | ✔ |
| HR Variability | ✔ | | | | | ✔ | ✔ |
| Sleep Variance | ✔ | | | | | ✔ | ✔ |
| Mobility Variance | ✔ | | | | | | |
| Psychomotor | ✔ | | | | | | |
| Respiratory Acoustics | ✔ | | | | | ✔ | |

## Participant Recruitment Process

Participants are recruited using different methods including through clinical services, hospitals, remotely including through citizen science approach, depending on the study criteria. A number of recruitment strategies are supported by the platform including:

- All participants are recruited at once and study starts simultaneously.
- Participants enter the study in a "batch" mode.
- Participants are recruited continuously until the desired sample size or date is reached ("stream mode").

Simultaneous recruitment from multiple sites is possible supporting recruitment of diverse population groups for the same study.

## Projects using RADAR-base platform

Table 3 presents a summary of selected projects using the platform with disorders they are focused on, along with the cohort size and sensors being used. It also lists the main objectives of the project. Brief summary of the projects listed in Table 3 are provided in the next sections. In some projects wearables data collected through the platform is augmenting existing collected data from different cohorts.

Table 3: RADAR-base platform projects summary.

| Project | Disease Area | Size | Main Devices/Data Types | Main Objectives |
|---|---|---|---|---|
| RADAR-CNS | Depression, | 600 | Fitbit, Phone Sensors, Questionnaires | Depressive Relapse |
| | Epilepsy, | 200 | Biovotion Everion, Empatica E4, Questionnaires | Epilepsy seizure and pre-ictal seizure detection |
| | Multiple Sclerosis | 500 | Fitbit, Bittium Faros, Phone Sensors, Questionnaires | Trajectory of disease, characterisation relapsing/remitting of disease symptoms |
| ART-CARMA | Cardiometabolic Risk Factors | 300 | Empatica EmbracePlus, Phone Sensors, Questionnaires | Pre-treatment initiation through to treatment initiation, titration and the subsequent period |
| ART | Attention Deficit Hyperactivity Disorder | 25 | Fitbit, Phone Sensors, Questionnaires | To establish a remote assessment and monitoring system for adults and adolescents with ADHD |
| RADAR-AD | Alzheimer's Disease | 200 | Fitbit, Phone Sensors, Questionnaires | Feasibility of remote monitoring technologies for AD |
| AIMS-2-TRIALS | Autism | 500 | Empatica E4, Fitbit, Phone Sensors, Questionnaires | Biology of autism to tailor treatments and develop new therapies and medicines. |
| BigData@Heart | Atrial Fibrillation | 160 | Phone Sensors, Questionnaires | Comparison of two strategies of rate-control, based either on initial treatment with digoxin or beta-blockers |
| DynaMORE | Mental Health | | Phone Sensors, Questionnaires | Developing an in silico model of stress resilience |
| CONVALESCENCE | Long-term Effects of COVID-19 | 800 | Garmin Vivoactive, Phone Sensors, Questionnaires | Characterisation, determinants, mechanisms and consequences of the long-term effects of COVID-19 |
| COVID-Collab | COVID-19 | 15000 | Fitbit, Questionnaires | Behaviour and physical and mental health occur in response to COVID infection, persistent symptoms, and the pandemic in general |
| RALPMH | Lung Disease | | Garmin Vivoactive, Phone Sensors, Questionnaires | Feasibility of remote monitoring technologies for high-burden pulmonary disorders |
| EDIFY | Eating Disorder | 500 | Oura Ring, Phone Sensors, Questionnaires | Delineating illness and recovery trajectories to inform personalised prevention and early intervention in young people |
| UNFOLD | Psychosis | 50 | Questionnaires | To characterise the processes involved in developing an identity as a person in recovery |
| Jovens na Pandemia & MAAY Study | Depression | 280 | Phone Sensors, Questionnaires | Remotely monitor behavioural and symptom changes associated with behavioural interventions in children and adolescents. |

### Remote Assessment of Disease And Relapse – Central Nervous System (RADAR-CNS)

RADAR-CNS was a cohort study that developed new ways of monitoring major depressive disorder, epilepsy, and multiple sclerosis using wearable devices and smartphone technology. Patients' data were collected continuously for 24 months (26). More than 1200 participants took part in the study in different disease areas, participant recruitment was done via clinics and hospitals. Different study protocols with different wearable devices were used for each disease. Participants were recruited from 6 different sites from different countries.

Digital biomarkers developed through the remotely collected data give a better understanding of the diseases and will help clinicians to manage them timely (27) (28) (22).

### ADHD Remote Technology Study of Cardiometabolic Risk Factors and Medication Adherence (ART-CARMA)

ART-CARMA aims to obtain real-world data from the patients' daily life to explore the extent to which ADHD medication and physical activity, individually and jointly, may influence cardiometabolic risks in adults with ADHD. The second objective is to obtain valuable real-world data on adherence to pharmacological treatment and its predictors and correlates. The long term goal is to use collected data to improve the management of cardiometabolic disease in adults with ADHD, and to improve ADHD medication treatment adherence and the personalisation of treatment (29). For this cohort two study sites in London and Barcelona are concurrently recruiting the participants using the platform.

### ADHD Remote Technology (ART)

The ART was a pilot project focused on developing a novel remote assessment system for Attention Deficit Hyperactivity Disorder (ADHD). ART assessed the feasibility and validity of remote researcher-led administration and self-administration of modified versions of two cognitive tasks sensitive to ADHD, a four-choice reaction time task (Fast task) and a combined Continuous Performance Test/Go No-Go task (CPT/GNG) (30). Cohort of 40 participants were recruited, 20 controls and 20 patients with ADHD.

### Remote Assessment of Disease And Relapse – Alzheimer's Disease (RADAR-AD)

RADAR-AD aimed to transform Alzheimer's disease patient care through remote assessment using mobile technologies such as smartphones or fitness trackers (31). The project developed the technology to identify which clinical or physiological features, digital biomarkers, can be measured remotely to predict deterioration in function. RADAR-AD created a pipeline for developing, testing and implementing remote measurement technologies with patients involved at each stage. Complete details of the study protocol and pipeline development is explained in (32). It was an augmentation study in which 300 participants took part. Three different categories of participants were recruited, controls, mild cognitively impaired (MCI)/prodromal AD and Alzheimer's dementia.

### Autism Innovative Medicine Studies-2-Trials (AIMS-2-TRIALS)

The AIMS-2-TRIALS programme includes a range of studies to explore how autism develops, from before birth to adulthood, and how this varies in different people. AIMS-2-TRIALS is looking for biological markers which indicate whether a person has or may develop particular characteristics (33). AIMS-2-TRIALS is collecting both active and passive data in clinical assessment settings and in home-based and ambulatory settings. Fitbit is used for remote data collection and Empatica E4 is used for local data collection at hospitals. Digital markers could help to identify those who may ultimately benefit from particular treatments. Medicines will also be tested to help with social difficulties, repetitive behaviours and sensory processing. Remote monitoring data is augmenting the clinical data.

### Rate Control Therapy Evaluation in Permanent Atrial Fibrillation (BigData@Heart)

RATE-AF study was designed to compare two strategies of rate-control, based either on initial treatment with digoxin or beta-blockers in 160 patients with Atrial fibrillation (AF) in need for rate control therapy. Monitoring with wearable devices, phone sensors and questionnaires was conducted over a continuous 6-month period. Objectives of the project included discovering new phenotypes, developing reliable sub-phenotyping and informing new taxonomies of heart failure based on a better understanding of underlying disease processes. Sleep, Heath Rate, Heart Rate Variation, and Activity data were collected to develop new phenotypes. This work is additionally supported by the EU IMI2 BigData@Heart major programme (34).

### Dynamic Modelling of Resilience (DynaMORE)

DynaMORE generated the first personalised in silico model of mental health in the face of adversity, or stress resilience. The model is based on and validated against unique multiscale longitudinal real-world empirical data sets, collected through neuroimaging, experimental assessments, questionnaires and remote monitoring using the pRMT app and a wearable device. The model will substantially deepen scientific understanding of the mechanisms of resilience, supporting the creation of mechanistically targeted interventions for the primary prevention of stress-related disorders. On this basis, DynaMORE developed an entirely new mobile health (mHealth) product incorporating the RADAR-base platform that will include model-based prognostic tools for real-time and real-life monitoring of at-risk subjects and for automated decision-making about timed, personalised interventions.

### CONVALESCENCE

CONVALESCENE is focused on the characterisation, determinants, mechanisms and consequences of the long-term effects of COVID-19, providing the evidence base for health care services (35). It's an existing large longitudinal cohort being further characterised and augmented by incorporating wearables data. Deep phenotyping and remote assessment using mobile devices and smartphones through RADAR-bae platform is being used to identify subclinical damage or dysfunction in individuals with long-term COVID-19.

### COVID Collab

COVID-Collab is a citizen science project with members of the public volunteering to donate their wearable data and complete diagnosis and symptom questionnaires. The main aim was to investigate the ongoing COVID-19 outbreak - 1) establish if wearable data can be used to diagnose COVID infection 2) characterise the disease symptoms and evolution. A key feature of the study is the use of wearables data to investigate changes in mental health and physiological measurements such as heart rate during infection with coronavirus (36).

### RALPMH: Remote Assessment of Lung Disease and Impact on Physical and Mental Health

Chronic lung disorders like chronic obstructive pulmonary disease (COPD) and idiopathic pulmonary fibrosis (IPF) are characterised by exacerbations and decline over time. 20 participants were recruited in each of three cohorts (COPD, IPF, and post-hospitalisation COVID). Data collection is being done remotely using the RADAR-base platform for different devices, including Garmin wearable devices and smart spirometers, mobile app questionnaires, surveys, and finger pulse oximeters. The RALPMH project is studying the feasibility of remote monitoring in chronic lung disorders and provide a reference infrastructure for future studies (37). Figure 5 shows an overview of the RALPHM study.

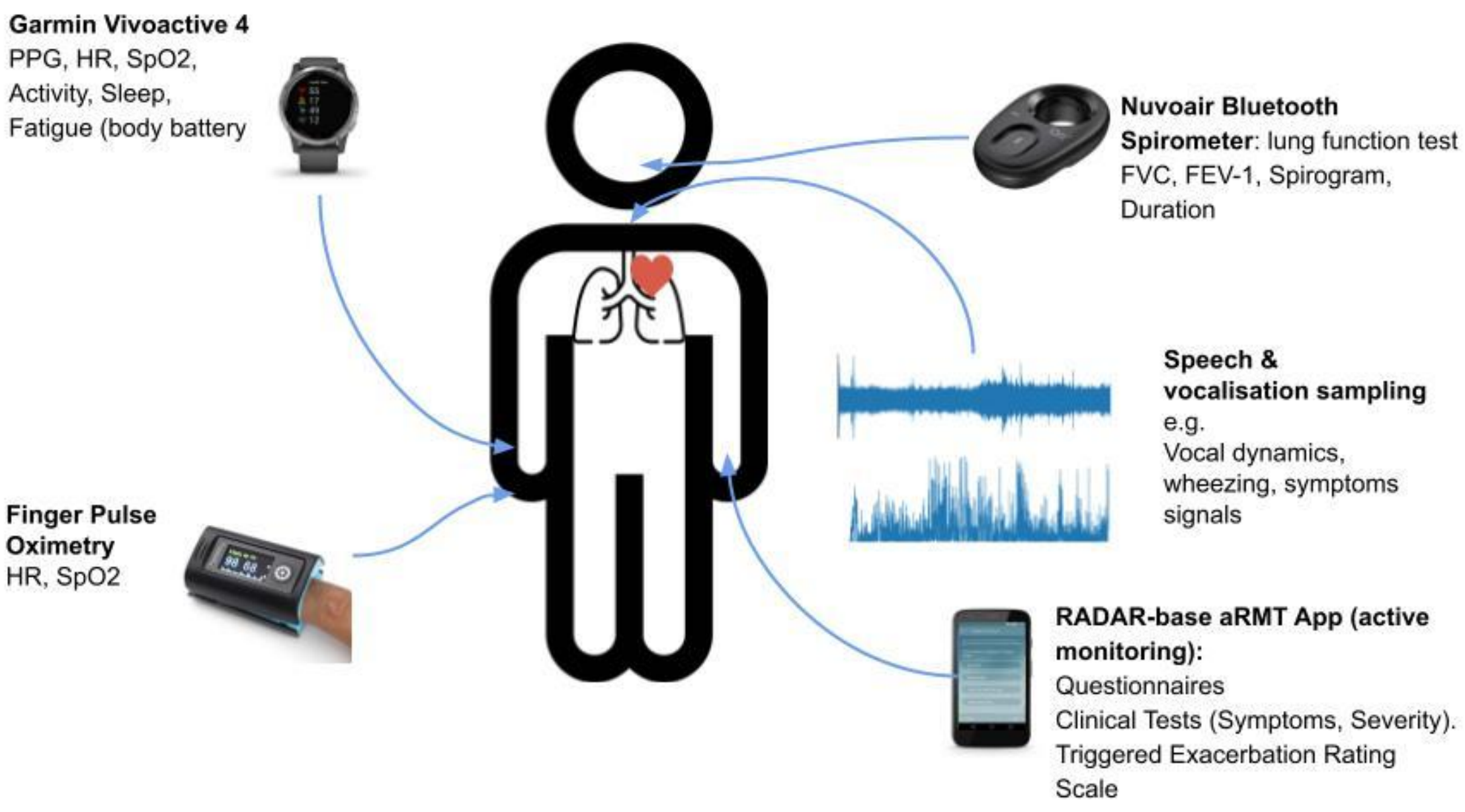

Figure 5. Various data sources will be used to collect both active and passive data to gain a unique perspective into patient health in the RALPMH lung disorder study.

### Eating Disorder (EDIFY)

The objective of the EDIFY study is to undertake a longitudinal comparison of the biopsychosocial symptom profiles of those with early and late-stage Eating Disorders(EDs), and recovery trajectories of those with early-stage EDs. This will provide evidence that will help inform decision-making of targeted intervention and preventative treatments across EDs for those with early and more progressed forms of illness. ED patients do not like to wear any wearable device or have any information displayed on their phone regarding their calories and daily workout. The Oura ring is being experimented for the first time with ED patients at scale as it provides no feedback to the patients during the study.

## Starting a Study with RADAR-base

RADAR-base is of interest to a wide range of mHealth communities from academic research to industry and sensor/wearable vendors interested in collecting data remotely or integrating new data sources into the platform. The platform is freely available as an open-source (Apache 2) GitHub repository (38) more details of the platform can be found on the official RADAR-base website (39). A detailed quickstart, deployment details and developer documentation are made available on the platform Confluence Wiki (40). Docker images for all the components are available at Docker Hub (41) and a Kubernetes stack is also available for the deployment (42). aRMT app questionnaires and protocol implementation is explained in (43). An exemplar Data Protection Impact Assessment (DPIA) is shared in Appendix 1.

## Results and Discussion

A Large number of mental and physical health research studies have employed the RADAR-Base platform for remote data collection with funding from many major funding agencies. This includes over 50 use cases exploring more than 30 disorder areas with more than 150,000 participants enrolled to date. Major disease areas using the platform are Major Depressive Disorder, Eating Disorder, Multiple Sclerosis, ADHD, Autism, Epilepsy, Atrial Fibrillation, Alzheimer's Disease, and COVID-19. Projects using the platform are collecting various health parameters depending on the disease area requirement. Data collected relates to cognition, mood, voice, digital usage, geolocation, heart rate, to name a few. Figure 6 and 7 shows examples of the status of collected data, its compliance and quality for different studies. Numerous challenges addressed by the platform include completeness of data, quality/accuracy of data, participant engagement and remote data collection.

The RADAR-base platform has effectively transformed low-level sensor data into digital biomarkers through feature generation pipelines. This process involves extracting relevant characteristics and patterns from the raw data, enabling the creation of meaningful and actionable insights. These digital biomarkers hold immense potential in various disease areas, aiding clinicians in making informed decisions, facilitating early intervention, and contributing to the prevention of relapse.

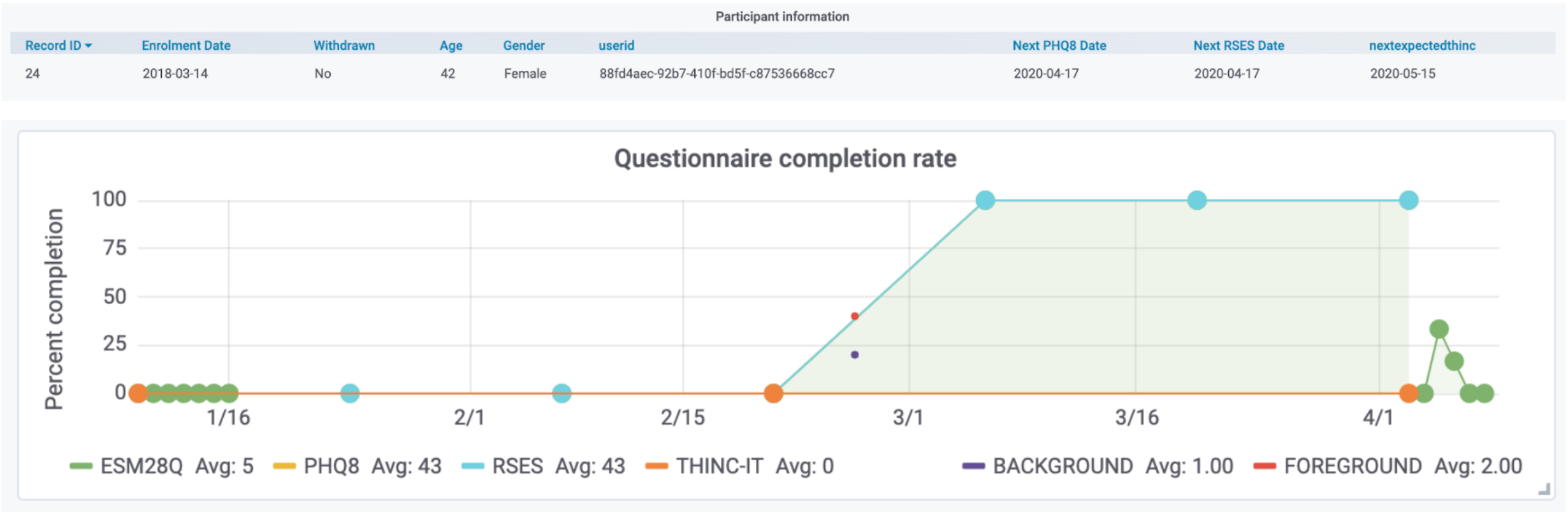

Figure 6. aRMT app questionnaire completion rate from a single patient from the RADAR-CNS Major Depression Study.

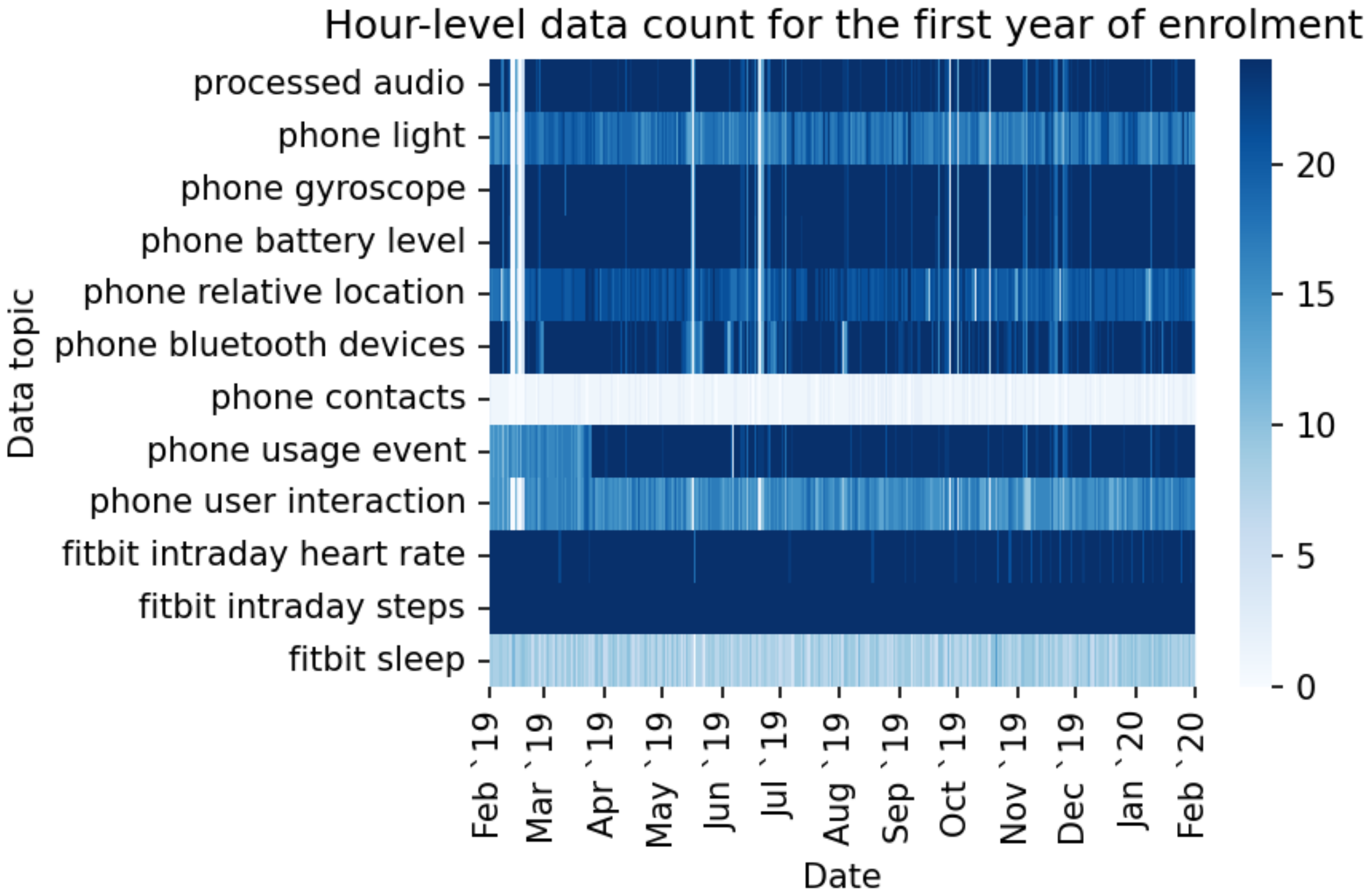

Figure 7. Contiguity of phone sensor data over the first year of enrolment collected through RADAR-base for a patient in the RADAR-CNS major depressive disorder study. The intensity of the colour represents how many hours in a day a particular modality is present.

The  key considerations for the development and evaluation of digital biomarkers are:

- Validation and Reliability: Ensuring that the digital biomarkers are accurate, reliable, and consistent in their measurements is crucial for their acceptance and adoption in clinical settings.

- Clinical Relevance: Assessing whether the digital biomarkers provide meaningful and clinically relevant information that can aid in disease diagnosis, treatment, or monitoring is essential.

- Data Privacy and Security: Handling sensitive health data from individuals requires robust privacy and security measures to protect patient information.

- Regulatory Compliance: Adhering to relevant regulatory guidelines and requirements is necessary to ensure that digital biomarkers meet the necessary standards for clinical use.

- Data Interoperability: Ensuring that digital biomarkers and the associated ecosystem can seamlessly integrate with existing healthcare systems and technologies is vital for their widespread adoption.

RADAR-base is a digital biomarker ecosystem that addresses these considerations. It provides a platform that focuses on safety and effectiveness by ensuring transparency in the algorithms used to generate biomarkers. Additionally, the interoperable components with open interfaces in RADAR-base facilitate the development of new multicomponent systems. This interoperability ensures that various digital biomarker sources can be integrated into a comprehensive and cohesive platform for healthcare purposes. Moreover, RADAR-base emphasises high integrity measurement systems, which means that the data collected and the biomarkers generated are reliable and accurate. This emphasis on data quality is essential for building trust in the digital biomarker ecosystem and encouraging its adoption in clinical research and routine patient care.

Overall, the systematic approach and emphasis on safety, effectiveness, transparency, and interoperability offered by RADAR-base can contribute to the advancement of digital biomarkers and their integration into healthcare systems, ultimately benefiting patient outcomes and medical research.

## Acknowledgements

This study has received support from the EU/EFPIA IMI Joint Undertaking 2 (RADAR-CNS grant No 115902). This communication reflects the views of the RADAR-CNS consortium, and neither IMI nor the European Union and EFPIA are liable for any use that may be made of the information contained herein. The authors receive funding support from the NIHR Biomedical Research Centre at South London and Maudsley NHS Foundation Trust and King's College London. The views expressed are those of the authors and not necessarily those of the NHS, the NIHR, or the Department of Health. The authors also acknowledge the support of NIHR University College London Hospitals Biomedical Research Centre.